\documentclass[acus]{JAC2000_mc} 
\usepackage{graphicx}
\begin{document}

\thispagestyle{empty}

\onecolumn

\begin{flushright}
{\large
SLAC--PUB--8870\\
June 2001\\}
\end{flushright}

\vspace{.8cm}

\begin{center}

{\LARGE\bf
Study of Low $\mathbf{\beta_y}$ Straight Section in SPEAR~3~\footnote
{\normalsize
{Work supported by Department of Energy contract  DE--AC03--76SF00515.}}}

\vspace{1cm}

\large{
Y.~Nosochkov, J.~Corbett and T. Rabedeau\\
Stanford Linear Accelerator Center, Stanford University,
Stanford, CA 94309}

\end{center}

\vfill

\begin{center}
{\LARGE\bf
Abstract }
\end{center}

\begin{quote}
\large{
The SPEAR~3 light source~\cite{spear3} has two 7.6~m straight sections one
of which is available for new insertion devices (ID).  By reducing the
vertical $\beta_y$ function at the center of the straight from 9.8~m to
1.5~m, the beam size can be decreased from 44~$\mu$m to 17~$\mu$m to reduce
the ID gap, but the ID length may be limited by the increased $\beta_y$
variation in the ID.  Alternatively, a ``double waist'' optics with two
symmetric low $\beta_y$ locations and a chicane may be considered to
implement two IDs.  In this paper, we discuss the low $\beta_y$ optics,
effects on the beam dynamic aperture, and the ID parameters needed to
maximize photon flux density and brightness while maintaining electron beam
lifetime.
}
\end{quote}

\vfill

\begin{center}
\large{
{\it Presented at the 2001 Particle Accelerator Conference
(PAC 2001)\\
Chicago, Illinois, June 18--22, 2001}
} \\
\end{center}

\newpage

\pagenumbering{arabic}
\pagestyle{plain}

\twocolumn

\title
{STUDY OF LOW $\beta_y$ STRAIGHT SECTION IN SPEAR~3~\thanks 
{Work sponsored in part by DOE Contract DE-AC03-76SF00515 and the Office 
of Basic Energy Sciences, Division of Chemical Sciences.}\vspace{-3mm} }

\author{Y.~Nosochkov, J.~Corbett and T. Rabedeau\\ 
SLAC, Stanford University, Stanford, CA 94309, USA}

\maketitle

\begin{abstract} 

The SPEAR~3 light source~\cite{spear3} has two 7.6~m straight sections one
of which is available for new insertion devices (ID).  By reducing the
vertical $\beta_y$ function at the center of the straight from 9.8~m to
1.5~m, the beam size can be decreased from 44~$\mu$m to 17~$\mu$m to reduce
the ID gap, but the ID length may be limited by the increased $\beta_y$
variation in the ID.  Alternatively, a ``double waist'' optics with two
symmetric low $\beta_y$ locations and a chicane may be considered to
implement two IDs.  In this paper, we discuss the low $\beta_y$ optics,
effects on the beam dynamic aperture, and the ID parameters needed to
maximize photon flux density and brightness while maintaining electron beam
lifetime.

\end{abstract}

\vspace{-1mm} 
\section{INTRODUCTION}

The SPEAR~3 design optics consists of two periodic arcs separated by the
East and West matching cells~\cite{lattice}.  Intended locations for the
insertion devices in the arcs are the twelve 3.1~m drift sections between
the 14 arc cells and four 4.8~m drifts at the arc ends.  At present, the
matching cells (MC) have identical symmetric optics with a 7.6~m drift at
the center of MC and phase advance $\mu_x/\mu_y\!=\!1.6/0.8$ [$2\pi$] per
cell.  The East 7.6~m drift is available for the new IDs, and the West MC
will contain the RF accelerating cavities.

The photon flux density and brightness in the insertion devices depend on
the electron beam size and, hence, $\beta$ functions at the IDs.  At
present, the $\beta_x/\beta_y$ values are $10.2/4.7$~m at the arc drift
sections, and $5.0/9.8$~m at the center of 7.6~m drifts.  For the SPEAR~3
design vertical emittance of $0.2$~nm at 3~GeV, the vertical beam size is
31 and 44~$\mu$m at the ID locations in the arcs and MCs, respectively.
The beam size and the ID gap can be reduced by decreasing the $\beta_y$ at
ID, but this will increase the beam divergence and variation of $\beta_y$
in the ID.  For a better ID performance, a large $\beta_y$ variation in the
insertion device should be avoided by limiting the ID length or the minimum
value of $\beta_y$.  The $\beta_y$ value is also limited by the acceptable
size of dynamic aperture and beam lifetime.

In this report, we discuss the low $\beta_y$ options for the 7.6~m drift in
the East MC.  The West MC was maintained close to nominal in most of this
study.  The optics and dynamic aperture calculations were done using
MAD~\cite{mad} and LEGO~\cite{lego} codes, respectively.

The present SPEAR~3 optics has been well optimized for a maximum dynamic
aperture.  To maintain the current lattice properties, the following
requirements were used for the low beta modifications:
\begin{itemize}
\vspace{-2mm} 
\item Matched optics using local quadrupoles.  
\vspace{-2mm}
\item Minimal change of the local phase advance.  
\vspace{-2mm}
\item Minimal increase of maximum $\beta$ functions.  
\vspace{-2mm} 
\item Realistic quadrupole strengths.  
\vspace{-2mm} 
\end{itemize}

In addition, we tried to keep equal or close phase advance in the East and
West MCs and limit the sextupole strengths, since the stronger sextupoles
and ring optical asymmetry may enhance resonance effects and reduce
dynamic aperture.

\begin{figure}[tb]
\centering
\includegraphics*[width=55mm, angle=-90]{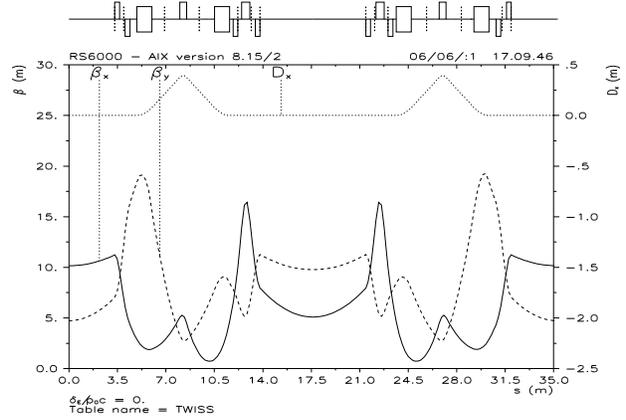}
\vspace{-1mm}
\caption{Nominal lattice functions in the MC.}
\label{opt-nominal}
\vspace{-3mm}
\end{figure}

\begin{figure}[b]
\centering
\includegraphics*[width=70mm]{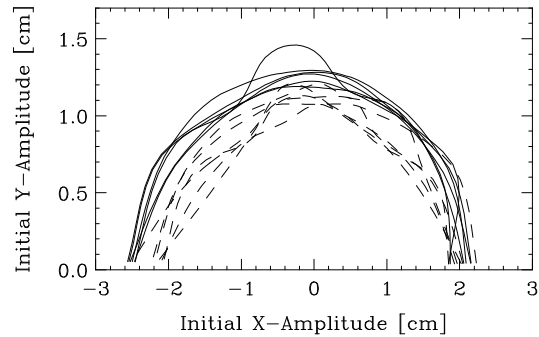}
\vspace{-3mm}
\caption{Nominal dynamic aperture for 6 error settings.}
\label{aper-nominal}
\vspace{-1mm}
\end{figure}

The nominal lattice functions in the MC are shown in
Fig.~\ref{opt-nominal}, where the 7.6~m drift is at center.  The
corresponding dynamic aperture without ID effects is shown in
Fig.~\ref{aper-nominal}, where the solid and dash lines are for the nominal
and 3\% off-energy particles, respectively.  The dynamic aperture
simulations included a full set of magnet errors and machine correction for
six random error settings.

\vspace{-1mm} 
\section{SINGLE LOW $\beta_y$ IN THE MC}

Several options for a matching cell with single low $\beta_y$ were analyzed
as listed in Table~1, where $\beta_{x,y}$ is at the MC center, $\mu_{x,y}$
is the MC phase advance, $\xi_{x,y}$ the ring natural chromaticity, and
$K_{SF,SD}$ the K-values of global sextupoles.

\begin{table}[tb]
\vspace{-3mm} 
\begin{center}
\caption{Single low $\beta_y$ options in the East MC.} 
\vspace{2mm} 
\begin{tabular}{|l|c|c|c|c|} 
\hline
& \multicolumn{2}{c|}{$\beta_y/\beta_x$ [m]} & 
  \multicolumn{2}{c|}{$\mu_x/\mu_y$ $[2\pi]$} \\
\cline{2-5} 
& East & West & East & West \\ 
\hline 
1 & 1.0 / 4.41 & 9.78 / 5.08 & 1.60 / 0.90 & 1.60 / 0.80 \\
2 & 1.0 / 3.77 & 9.78 / 6.49 & 1.56 / 0.96 & 1.56 / 0.96 \\
3 & 1.0 / 4.35 & 9.77 / 5.04 & 1.60 / 0.90 & 1.60 / 0.90 \\
4 & 1.5 / 5.25 & 9.00 / 5.09 & 1.60 / 0.87 & 1.60 / 0.87 \\
5 & 2.0 / 6.04 & 9.17 / 5.09 & 1.60 / 0.84 & 1.60 / 0.84 \\
\hline
\end{tabular}
\\[1mm]
\begin{tabular}{|l|c|c|}
\hline
& $\xi_x/\xi_y$ & $K_{SF}/K_{SD}$ $[m^{-3}]$ \\
\hline
1 & -22.20 / -15.35 & 31.43 / -37.96 \\
2 & -22.67 / -15.70 & 31.21 / -37.30 \\
3 & -22.26 / -15.51 & 31.46 / -37.79 \\
4 & -22.03 / -15.19 & 31.09 / -37.41 \\
5 & -21.90 / -15.03 & 30.85 / -37.26 \\
\hline
\end{tabular} 
\end{center} 
\vspace{-6mm} 
\end{table}

The low $\beta_y$ optics in the East MC was matched using five local
quadrupole families while the West MC was maintained close to the nominal.
For low $\beta$ peaks and chromaticity, the MC phase advance was slightly
varied.  The arc phase was adjusted accordingly to keep the working tune at
$\nu_x/\nu_y\!=\!14.19/5.23$.

Comparison of the three $\beta_y\!=\!1$~m options showed a larger dynamic
aperture in case 3.  It indicates that the preferred optical conditions are
close to the nominal phase advance, equal phase advance in the two MCs,
low chromaticity and weak sextupoles. Note that low $\beta_y$ naturally
increases the MC vertical phase advance since 
$\mu\!=\!\int \frac{ds}{\beta}$.

Cases 4,5 with $\beta_y\!=\!1.5$ and 2~m were designed similar to case 3.
In comparison, the lower $\beta_y$ requires stronger focusing and larger
quadrupole strengths as listed in Table~2.  Consequently, the lower
$\beta_y$ leads to larger $\beta$ peaks in the MC, higher chromaticity and
stronger sextupoles, and reduced dynamic aperture.  At present, the QDXE,
QFXE design strengths are limited at $|K|\!\approx\!2$, therefore optics
with $\beta_y\!<\!1$~m would require an upgrade of these magnets.

\begin{table}[b]
\vspace{-6mm} 
\begin{center}
\caption{Quadrupole K-values in the East MC $[m^{-2}]$.} 
\vspace{2mm} 
\begin{tabular}{|l|rrr|} 
\hline
$\beta_y$ [m] & 1.0 (c.3) & 1.5 & 2.0 \\
\hline
QDXE & -1.9269 & -1.6027 & -1.3254 \\
QFXE &  1.8308 &  1.7695 &  1.6880 \\
QDYE & -0.8003 & -0.9760 & -1.0481 \\
QDZE & -0.6667 & -0.7032 & -0.7245 \\
QFZE &  1.3458 &  1.4110 &  1.4512 \\
\hline
\end{tabular} 
\end{center} 
\vspace{-4mm} 
\end{table}

\begin{figure}[tb]
\centering
\includegraphics*[width=55mm, angle=-90]{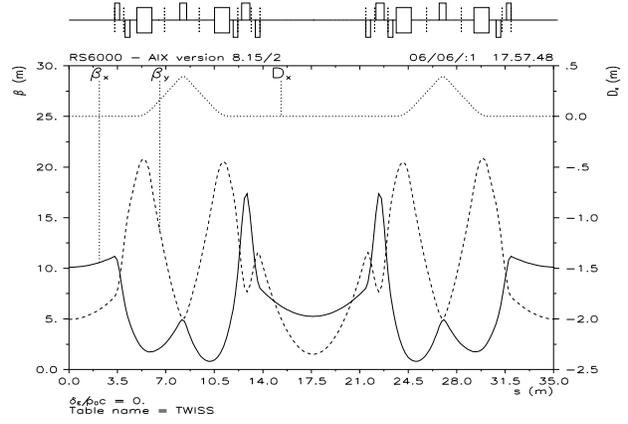}
\vspace{-1mm}
\caption{Matching cell with a single $\beta_y\!=\!1.5$~m .}
\label{opt-single}
\vspace{-0mm}
\end{figure}

\begin{figure}[tb]
\centering
\includegraphics*[width=70mm]{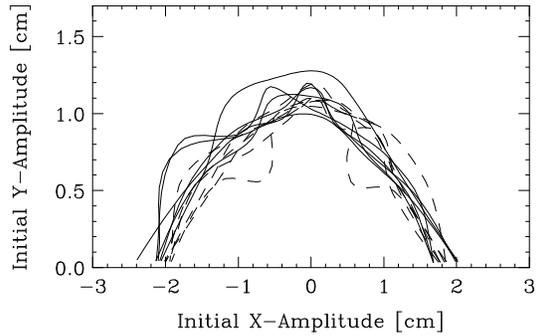}
\vspace{-3mm}
\caption{Dynamic aperture for a single $\beta_y\!=\!1.5$~m.}
\label{aper-single}
\vspace{-4mm}
\end{figure}

Maintaining sufficient aperture for horizontal injection and beam lifetime
with Touschek effects requires a large horizontal dynamic aperture with up
to 3\% energy errors.  LEGO simulations with machine errors, but without ID
effects, estimate the horizontal dynamic aperture near 16~mm for
$\beta_y\!=\!1$~m, and 17~mm for $\beta_y\!=\!1.5$ and 2~m, compared to the
nominal 18-20~mm aperture.  This aperture may be sufficient even with the
ID effects included which will reduce it by $\sim$1-2~mm.  For a
conservative design, we consider an option with $\beta_y\!=\!1.5$~m.  The
MC optics and dynamic aperture in this case are shown in
Fig.~\ref{opt-single} and~\ref{aper-single}.

It had been shown that the SPEAR~3 horizontal emittance can be reduced by
allowing a small dispersion $\eta_x\!\sim\!10$~cm in the arc
drifts~\cite{safranek}.  The required dispersion was obtained by slightly
mismatching the arc achromats and appropriate adjustment of the matching
cells.  In this low $\beta_y$ study, we verified dynamic aperture for
$\eta_x\!=\!5$ and 10~cm in the arc drifts.  LEGO simulations without ID
effects showed that at $\beta_y\!=\!1.5$~m the horizontal dynamic aperture
is 15-16~mm at $\eta_x\!=\!5$~cm, and $\sim$13~mm at 10~cm dispersion.  We
conclude that dispersion up to 5~cm may be used to reduce the horizontal
emittance from 18.5 to 14.6~nm.

The low $\beta_y$ function increases beam divergence and $\beta_y$
variation in the ID and may limit the ID length.  At distance $s$ from
the waist:  $\beta_{y}(s)\!=\!\beta_y\!+\!\frac{s^2}{\beta_y}$, where
$\beta_y$ is at the waist.  For example, for $\beta_{y}(s)$ variation below
100\% and the waist at ID center, the half ID length has to be less than
$\beta_y$.

\vspace{-1mm} 
\section{DOUBLE LOW $\beta_y$ IN THE MC}

\begin{figure}[tb]
\centering
\includegraphics*[width=55mm, angle=-90]{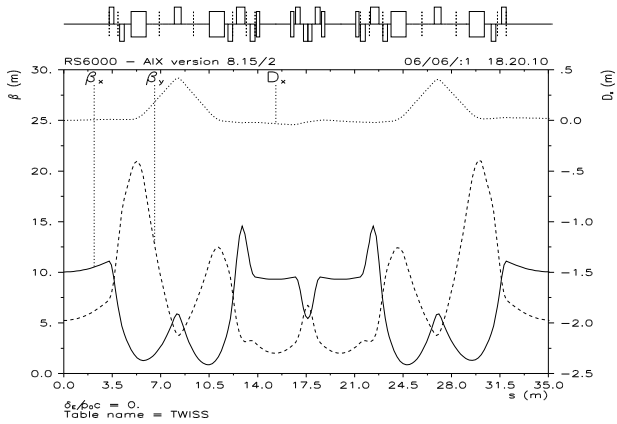}
\vspace{-1mm}
\caption{Matching cell with a double $\beta_y\!=\!2$~m .}
\label{opt-double}
\vspace{-3mm}
\end{figure}

The 7.6~m straight section in the matching cell can be upgraded to have two
low $\beta_y$ locations for two IDs.  A double waist optics can be obtained
using an extra quadrupole triplet at center of the straight.  An example of
such optics with two symmetric $\beta_y\!=\!2$~m locations is shown in
Fig.~\ref{opt-double}, where the space available for each ID is about 2~m
and the six local quadrupole families are adjusted to keep the MC matched.

The separation of the two ID beam lines can be done using a four bend
horizontal chicane.  We investigated symmetric and anti-symmetric chicane
schemes with 25~cm dipoles placed symmetrically around the $\beta_y$
waists as shown in Fig.~\ref{opt-double}.  In the symmetric scheme, the
bending angles in the four dipoles were [-10,+10,+10,-10]~mrad which
produce 24.7~mm closed orbit bump and 20~mrad angle between the beam lines.
In the anti-symmetric scheme, the bending angles were
[-8,+25.7,-25.7,+8]~mrad to generate two $\pm19.8$~mm bumps and parallel ID
beam lines with horizontal separation of 57~mm.

The chicane bends generate additional dispersion in the MC.  We found that
with the symmetric chicane this dispersion can be locally compensated by
slightly mismatching the MC achromats.  In the anti-symmetric scheme,
the dispersion could not be fully compensated using local quadrupoles, but
the residual dispersion in the ring is below 4~cm which is not a problem
for the IDs.  This dispersion has only minor reduction effect on the
emittance.  The anti-symmetric double waist optics is shown in
Fig.~\ref{opt-double}.

The double $\beta_y$ optics further increases the vertical phase advance in
the MC.  Similar to the single $\beta_y$ optics, the $\mu_y$ has to be kept
close to the nominal value for a better dynamic aperture.  For realistic
magnet strengths and matched optics, the vertical phase advance was set at
1.1~[$2\pi$] as compared to the nominal 0.8~[$2\pi$].  The MC sextupole
strengths were optimized for larger dynamic aperture.

\begin{figure}[tb]
\centering
\includegraphics*[width=70mm]{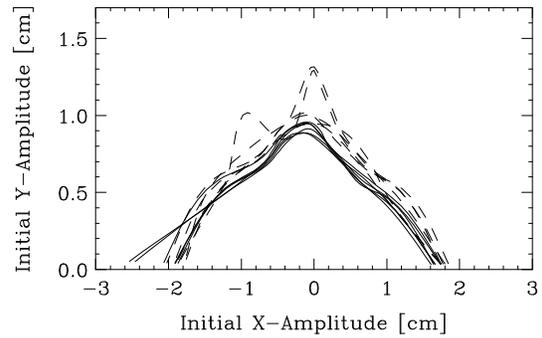}
\vspace{-3mm}
\caption{Dynamic aperture for a double $\beta_y\!=\!2$~m.}
\label{aper-double}
\vspace{-3mm}
\end{figure}

LEGO simulations showed a larger dynamic aperture in the anti-symmetric
scheme as shown in Fig.~\ref{aper-double}.  This aperture may be sufficient
for the beam operation, but more study is required for the design of ID
beam lines.

\vspace{-1mm}
\section{UNDULATOR PERFORMANCE}

\begin{figure}[htb]
\centering
\includegraphics*[width=65mm]{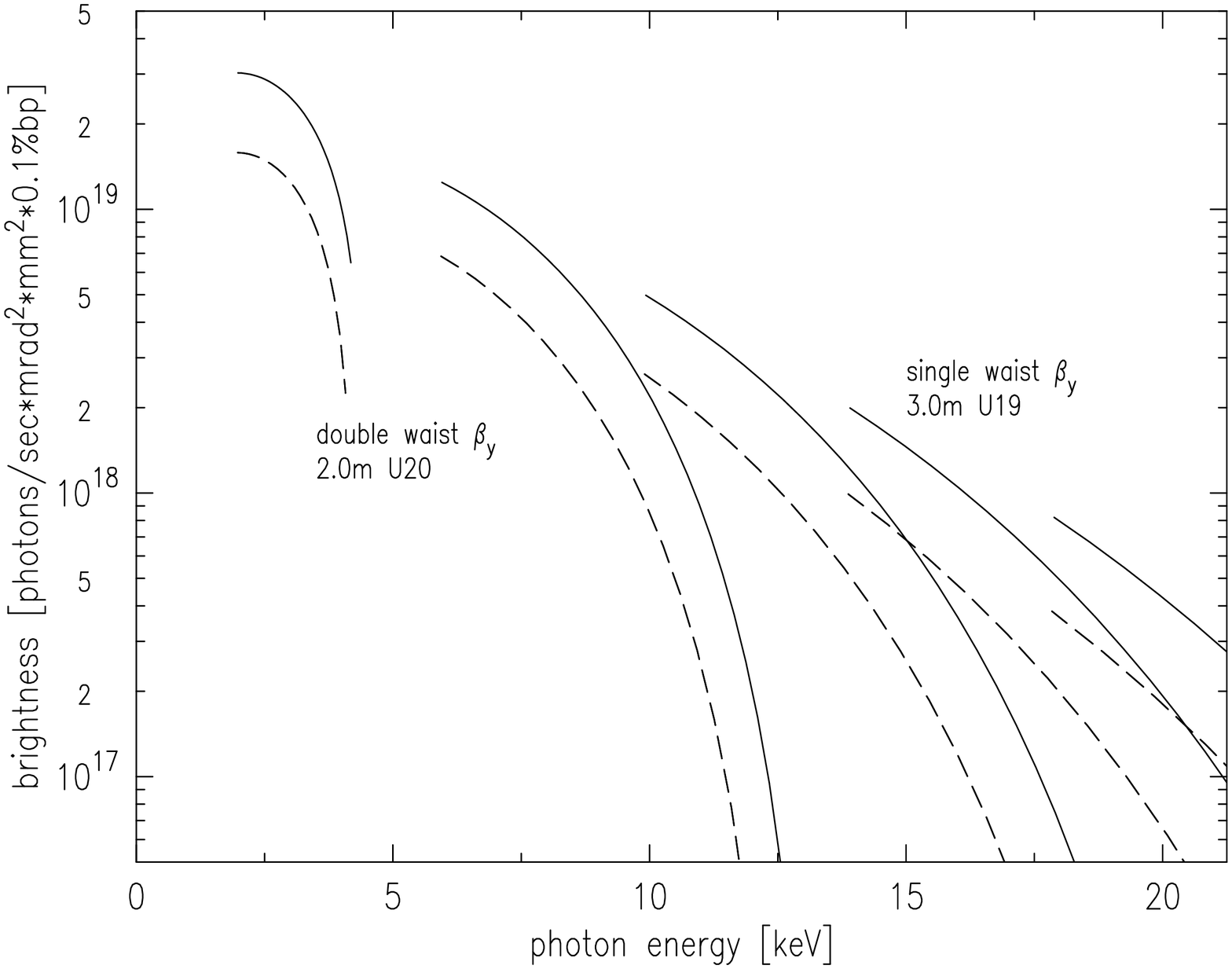}
\vspace{-3mm}
\caption{Undulator tuning curves for the single (solid line) and double
(dash) waist low $\beta_y$ optics.}
\label{bright}
\vspace{-3mm}
\end{figure}

The primary motivation for reducing $\beta_y$ in the matching cells is to
enhance hard x-ray brightness from small gap undulators.  To evaluate the
potential performance of small gap undulators in the MC, a rough
examination of suitable undulator parameters was undertaken using the XTC
routine of the XOP software suite~\cite{xop}.  Tuning curve simulations for
hybrid, planar undulators in the absence of field errors are shown in
Fig.~\ref{bright} for the ID parameters listed in Table~3.  The curves
depict the fundamental and odd harmonics through the 9th order.  For the
high brightness hard x-ray beams, the undulator parameters were selected to
provide reasonable tuning curve overlap in the $E\!>\!6$~keV regime without
concern for the coverage gap between the first and third harmonics for
$E\!<\!6$~keV.  The magnet gaps used in the simulations preserve reasonable
beam lifetime, though somewhat improved performance at high energies could
be obtained using smaller gaps.

\begin{table}[htb]
\vspace{-3mm} 
\begin{center}
\caption{Parameters of simulated IDs.} 
\vspace{3mm}
\begin{tabular}{|lcccc|} 
\hline
waist & $\beta_x/\beta_y$ & gap & period & length \\
 & [m] & [mm] & [mm] & [m] \\
\hline
single & 5.25 / 1.5 & 5.0 & 19.0 & 3.0 \\
double &  8.0 / 2.0 & 6.0 & 20.0 & 2.0 \\
\hline
\end{tabular} 
\end{center} 
\vspace{-6mm} 
\end{table}

\vspace{-1mm}
\section{CONCLUSION}

We conclude that the SPEAR~3 design permits single or double low $\beta_y$
optics in the matching cell 7.6~m drift with $\beta_y$ as low as 1~m.
Dynamic aperture for this optics seems to be acceptable for beam
operation.


\vspace{-1mm} 
\begin{thebibliography}{9}

\bibitem{spear3} R.~Hettel {\it et al.}, ``Design of the SPEAR 3 Light
Source,'' EPAC 2000, Vienna, June 2000.
\vspace{-1mm} 
\bibitem{lattice} J.~Corbett {\it et al.}, SLAC--PUB--7882, July 1998.
\vspace{-1mm}
\bibitem{mad} The MAD Home Page, http://wwwslap.cern.ch/mad/.
\vspace{-1mm} 
\bibitem{lego} Y.~Cai {\it et al.}, SLAC--PUB--7642, August 1997.
\vspace{-1mm} 
\bibitem{safranek} J.~Safranek, ``Non-Zero Dispersion in the SPEAR~3 
Straights,'' February 1998, unpublished.
\vspace{-1mm}
\bibitem{xop} M.~Sanchez del Rio, {\it et al.}, {\it SPIE}, \textbf{3152},
p. 148 (1997).

\end{thebibliography}
\end{document}